\documentclass{KapProc} 

\normallatexbib

\let\lcitebracket[
\let\rcitebracket]

\renewcommand{\theequation}{\arabic{section}.\arabic{equation}}

\begin{document}

\articletitle[MICROSYSTEMS AS CORRELATION CARRIERS]{DECOHERENCE VERSUS 
   THE IDEA\-LI\-ZA\-TION OF
  MICROSYSTEMS AS CORRELATION CARRIERS BETWEEN MACROSYSTEMS}

\author{Ludovico Lanz and Bassano Vacchini}
\affil{Dipartimento di Fisica dell'Universit\`a di
Milano and INFN, Sezione di Milano\\
Via Celoria 16, I-20133, Milano, Italy}
\email{lanz@mi.infn.it, vacchini@mi.infn.it}

\author{Olaf Melsheimer}
\affil{Fachbereich Physik, Philipps--Universit\"at\\
Renthof 7, D-35032, Marburg, Germany}
\email{melsheim@mailer.uni-marburg.de}


\begin{abstract}
It is argued that the appropriate framework to describe a microsystem as a
correlation carrier between a source and a detector is non-equilibrium
statistical mechanics for the compound source-detector system. An attempt
is given to elucidate how this idealized notion of microsystem might arise
inside a field theoretical description of isolated macrosystems:
then decoherence appears as the natural limit of this idealization.
\end{abstract}
\section{Preeminent role of field theory}
\label{s1}
\setcounter{equation}{0}
Even if it is generally accepted that quantum field theory must be
used in high energy physics, questions on foundations of quantum
mechanics, description of measuring process and discussion of
decoherence are usually addressed to in the context 
of the N particle generalization of
the Schr\"odinger equation, while in that context quantum field theory is
often only appreciated as a more refined tool to accommodate relativity 
and to account for particlelike aspects of electromagnetism.
This is
deeply rooted in mechanics and in the atomistic picture of matter. 
However one runs into difficulties and puzzles: objective properties
for particles cannot be reconciled with quantum mechanics, quantum
mechanical models of the measuring process are hardly compatible with
objective description of macrosystems, decoherence must be supplied to 
the Schr\"odinger equation, either due to lack of isolation in any
system, or by some additional stochasticity. We stress the point
of view that the concept of a physical process running inside a
suitably prepared isolated system and displayed by a certain set of
relevant variables must be the starting point and that the theoretical 
description should be based on quantum field theory of finite
systems. This point of view is much closer to thermodynamics than to
mechanics: the basic ideas linked to atomistic structure of matter are 
however kept into account in a more subtle way by the quantization of
the fields underlying the physical model and by the locality or
quasi-locality of their interaction. The concept of a particle arises
only in an unsharp way when one is pursuing universal features arising 
from locality of field theory and polishing away what comes from
boundary conditions and residual interactions. In the most striking
way a particle emerges when a process can be performed in which a
\emph{source part} of a macrosystem affects a \emph{detecting part} of it
through a microchannel consisting of a one- or few-particle
system produced by the first part and directed to the second
one. Looking at the problem this way, decoherence is
obviously already inside the description: actually it is a hard
theoretical job to drive it back from the microchannel, completely in
tune with what experimentalists do. On the 
contrary the usual theoretical setting seems strange since it makes
theorists work putting decoherence in, while experimentalists work
hard to drive it back. This point of view
about particles goes back to Ludwig's approach to quantum mechanics.
By suitable axiomatization of general features of particular processes 
which give evidence of particles he succeeded in obtaining as a
description of these processes quantum mechanics already in the modern 
form~\cite{Ludwig} (p.o.v.\ measures, operations, instruments) that is
now generally recognized~\cite{modern} as the formalism adequate to
describe in a realistic way processes due to microsystems. Obviously
when highly idealized schematizations can be applied, typically if
decoherence can be neglected, and space-time symmetry for the
microsystem can be assumed, the more schematic Dirac's book axiomatics 
emerges in all its geometrical neatness. While Ludwig pointed to a new 
theory encompassing microsystems and macrosystems in order to set the
duality micro-macrosystems, we try to do this remaining inside
quantum field theory, only improving somehow non-equilibrium theory
for isolated systems. In \S~\ref{s2} we simply describe how a
microchannel can arise, in \S~\ref{s3} the general structure of
non-equilibrium theory is recalled and compatibility of
the general dynamics of the system with the presence of the
microchannel is indicated. The physical model we will use is
a self-interacting spinless Schr\"odinger field confined inside a
finite region $\omega$. It can be trivially improved using several
interacting fields with spin and should be amenable to the treatment
of bound states and resonances between them. However all this is a
very primitive stage since no intermediate gauge fields are
introduced. To the space region $\omega$ a set of \emph{normal modes}
$\{u_r (\textbf{x})\}$ is associated. They are an orthonormal,
complete set of solutions of the stationary state equation:
\begin{equation}
  \label{eq:1.1}
  -\frac{\hbar^2}{2m}\Delta_2 u_n (\textbf{x}) + {\cal V}(\textbf{x})u_n
  (\textbf{x})=W_n u_n (\textbf{x})
  \quad u_n (\textbf{x})=0, \> \textbf{x}\in \delta\omega, \> u_n\in
  L^2 (\omega);
\end{equation}
${\cal V}(\textbf{x})$ being a suitable potential for external and internal
effective forces. The Schr\"odinger field is defined by:
\begin{equation}
  \label{eq:1.2}
  {\hat \Phi} (\textbf{x})=\sum_{n} {\hat a}_n u_n({{\bf x}}), 
  \qquad
  [{\hat
    a}_n,{\hat a}^{\scriptscriptstyle \dagger}_{n'}]_\mp = \delta_{nn'}
\end{equation}
with ${\hat a}_n$ Bose or Fermi annihilation operators on the
Fock-space of the
system. The Hamiltonian of the system is assumed as:
\begin{equation}
  \label{eq:1.3}
  \hat H = \sum_{n} W_n {\hat a}^{\scriptscriptstyle \dagger}_{n}{\hat
    a}_n + \frac{1}{2} \int_\omega d^3\!\textbf{x} d^3\!\textbf{y}
  \,
  {\hat \Phi}^{\scriptscriptstyle \dagger} (\textbf{x}){\hat
    \Phi}^{\scriptscriptstyle \dagger} (\textbf{y}) V
  (|\textbf{x}-\textbf{y}|) {\hat \Phi} (\textbf{y}){\hat \Phi}
  (\textbf{x}), 
\end{equation}
$V(r)$ being the basic microphysical input, a short-range function
which gives the quasi-local form of interaction and will finally
represent two-body interaction between the particles of the system.
\section{The microchannel }
\label{s2}
\setcounter{equation}{0}
Postponing a more technical sketch of the treatment of a
non-equi\-li\-brium system, we now come to the main point: the 
microchannel. For this issue we choose a bundle of normal 
modes 
$
r \in M
$, $M$ being a suitable subset of the indexes $n$:
$M$ are the normal modes, $M^{\scriptscriptstyle C}$ the remaining
ones. The field operator
${\hat \Phi} (\textbf{x})$ contains them both
\begin{equation}
  \label{eq:2.1}
  {\hat \Phi} (\textbf{x}) = {\hat \Phi}_{M} (\textbf{x}) + {\hat
    \Phi}_{{M^{\scriptscriptstyle C}}} (\textbf{x}),  
 \end{equation}
\begin{displaymath}
    {\hat \Phi}_{M} (\textbf{x})=\sum_{r\in M} {\hat a}_r u_r({{\bf x}}),
  \qquad
  {\hat \Phi}_{{M^{\scriptscriptstyle C}}} (\textbf{x})=
  \sum_{s\in {M^{\scriptscriptstyle C}}} {\hat a}_s u_s({{\bf x}}).
\end{displaymath}
The idea of a microchannel is formalized assuming that during 
a time interval $[t_0,t_1]$ the channel modes are depleted, so that
the contribution to the dynamics of the system due to interaction
between channel-modes is negligible. Then there is a
possible dynamics of the system with unfeeded channel, described by a
statistical operator $\hat{\rho}_t^{\scriptscriptstyle 0}$ satisfying:
\begin{equation}
  \label{eq:2.2}
  {\hat a}_r \hat{\rho}_t^{\scriptscriptstyle 0} =0,
  \qquad
  \forall r \in M
\end{equation}
i.e., without excitations of $M$-modes and evolving according to
$
  {d\hat{\rho}_t^{\scriptscriptstyle 0}}/{dt}=
  -{i}/{\hbar}[\hat{H},\hat{\rho}_t^{\scriptscriptstyle 0}]
$.
There is however also a possible dynamics with feeded channel,
described by a statistical operator
\begin{equation}
  \label{eq:2.3}
  \hat{\rho}_t^{\scriptscriptstyle (1)}=\sum_{r,r'\in M} w_{rr'} (t)  {\hat
    a}^{\scriptscriptstyle \dagger}_r 
\hat{\rho}_t^{\scriptscriptstyle 0} {\hat a}_{r'},
\end{equation}
with one excitation related to $M$.
$\hat{\rho}_t^{\scriptscriptstyle (1)}$ 
is a positive operator if $w_{rr'} (t)$ is a
positive matrix and by (\ref{eq:2.2}) it is normalized if $\sum_r
w_{rr} (t)=1$. For $t\in [t_0,t_1]$ the following representation of
the statistical operator for a system endeavored with a 
microchannel should hold:
\begin{equation}
  \label{eq:2.4}
  \hat{\rho}_t = (1-\lambda)\, \hat{\rho}_t^{\scriptscriptstyle 0}  +
  \lambda \!\! 
  \sum_{r,r'\in M} w_{rr'} (t)  {\hat a}^{\scriptscriptstyle
    \dagger}_r 
  \hat{\rho}_t^{\scriptscriptstyle 0} {\hat a}_{r'} ,
  \qquad
  0<\lambda<1,
\end{equation}
$\lambda$ giving the probability that the microchannel is feeded. For
the statistical operator (\ref{eq:2.4}) by (\ref{eq:2.2}) the
interaction between modes in $M$ is negligible. We
assume at first that also the interaction between a mode $r\in
M$ and the modes $s\in {M^{\scriptscriptstyle C}}$ can be 
neglected at least in the time
interval $[t_0,t_1]$: then Liouville von Neumann equation for
$\hat{\rho}_t$ implies that
\begin{equation}
  \label{eq:2.5}
  \frac{dw_{rr'}}{dt} (t)=-\frac{i}{\hbar} (W_r - W_{r'})w_{rr'} (t).
\end{equation}
Eq.~(\ref{eq:2.5}) can be considered as the evolution equation of a
statistical operator $W^{\scriptscriptstyle (1)} (t)$ defined as
\begin{equation}
  \label{eq:2.6}
  W^{\scriptscriptstyle (1)} (t)= \sum_{r,r'\in M}|r\rangle w_{rr'}
  (t)\langle r'|, 
\end{equation}
describing the microsystem inside the microchannel:
\begin{displaymath}
  \frac{dW^{\scriptscriptstyle (1)} }{dt} (t)=-\frac{i}{\hbar}
  [H_0^{\scriptscriptstyle (1)},W^{\scriptscriptstyle (1)} (t)],  
  \qquad
  H_0^{\scriptscriptstyle (1)}=\sum_{r}|r\rangle W_{r} \langle r|,
\end{displaymath}
while $|r\rangle$ is a basis in the one-particle Hilbert space ${\cal
  H}^{\scriptscriptstyle M}\subset {L_2 (\omega)}$ spanned by $r\in M$,
$\langle\textbf{x}|r\rangle=u_r (\textbf{x})$. 
Taking an observable $\hat A$ of the system or more in particular an
element $\hat{E}^{\bf{\scriptscriptstyle A}} (S)$ of the spectral
measure of some 
commuting set of self-adjoint operators on some $\sigma$-algebra of
sets $S$, expectations or probability measures are given by
expressions:
\begin{equation}
  \label{eq:2.7}
  \textrm{Tr} (\hat A \hat{\rho}_t) 
  = 
  (1-\lambda)\textrm{Tr}(\hat A
  \hat{\rho}_t^{\scriptscriptstyle 0})+\lambda \!\! 
  \sum_{r,r'\in M} w_{rr'} (t)\textrm{Tr}( {\hat a}_{r'}
  \hat A {\hat a}^{\scriptscriptstyle \dagger}_r 
  \hat{\rho}_t^{\scriptscriptstyle 0} )
\end{equation}
\begin{displaymath}
   \textrm{Tr} (\hat{E}^{\bf{\scriptscriptstyle A}} (S) \hat{\rho}_t) 
  =
  (1-\lambda)\textrm{Tr}(\hat{E}^{\bf{\scriptscriptstyle A}} (S)
  \hat{\rho}_t^{\scriptscriptstyle 0})+\lambda \!\! 
  \sum_{r,r'\in M} w_{rr'} (t)\textrm{Tr}( {\hat a}_{r'}
  \hat{E}^{\bf{\scriptscriptstyle A}} (S) 
  {\hat a}^{\scriptscriptstyle \dagger}_r 
  \hat{\rho}_t^{\scriptscriptstyle 0} ).
\end{displaymath}
Setting $\textrm{Tr}( {\hat a}_{r'}
  \hat A {\hat a}^{\scriptscriptstyle \dagger}_r 
  \hat{\rho}_t^{\scriptscriptstyle 0} )
  =
  \langle r' | A_t^{\scriptscriptstyle (1)} | r\rangle$,
    $\textrm{Tr}( {\hat a}_{r'}
  \hat{E}^{\bf{\scriptscriptstyle A}} (S) 
  {\hat a}^{\scriptscriptstyle \dagger}_r 
  \hat{\rho}_t^{\scriptscriptstyle 0} )
  =
  \langle r' | F_t^{\scriptscriptstyle (1)} (S) | r\rangle$,
the r.h.s. of eq.~(\ref{eq:2.7}) related to the microchannel can be
written:
\begin{eqnarray}
  \label{eq:2.9}
  \sum_{r,r'\in M} w_{rr'} (t) \langle r' | A_t^{\scriptscriptstyle
    (1)} | r\rangle 
  &=&
  \textrm{Tr}_{{\cal
  H}^{\scriptscriptstyle M}} (W^{\scriptscriptstyle
    (1)}(t)A_t^{\scriptscriptstyle (1)}) 
  \\
  \sum_{r,r'\in M} w_{rr'} (t) \langle r' | F_t^{\scriptscriptstyle
    (1)} (S) | r\rangle 
  &=&
  \textrm{Tr}_{{\cal
  H}^{\scriptscriptstyle M}} (W^{\scriptscriptstyle
    (1)}(t)F_t^{\scriptscriptstyle (1)} (S)), 
  \nonumber
\end{eqnarray}
showing the typical mathematical structure of \emph{one-particle} quantum 
mechanics formulated in the Hilbert space ${\cal
  H}^{\scriptscriptstyle M}$. Let us
notice that even if $\hat{E}^{\bf{\scriptscriptstyle A}} (S)$ is a
Fock-space 
projection valued measure, the related
$\hat{F}^{\bf{\scriptscriptstyle A}} (S)$ is in general a  
p.o.v.\ normalized measure (positive operator valued or \emph{effect}
valued), according to modern axiomatics. 
Let us assume that $[\hat{E}^{\bf{\scriptscriptstyle A}},\hat{N}_{M}]$,
with $\hat{N}_{M}=\sum_{r\in M} {\hat a}^{\scriptscriptstyle \dagger}_r
{\hat a}_r$ and furthermore 
\begin{eqnarray*}  
  \lefteqn{\sum_{r} \langle r_1 | F_t^{\scriptscriptstyle (1)} (S) | r
    \rangle 
    \langle r | F_t^{\scriptscriptstyle (1)} (S) | r_2  \rangle  
    }
  \\
  &&  
  = \sum_{r}
  \textrm{Tr}
  ( {\hat a}_{r_1}
  \hat{E}^{\bf{\scriptscriptstyle A}} (S) {\hat a}^{\scriptscriptstyle
    \dagger}_{r}  
  \hat{\rho}_t^{\scriptscriptstyle 0} )
  \textrm{Tr}
  ( {\hat a}_{r}
  \hat{E}^{\bf{\scriptscriptstyle A}} (S) {\hat a}^{\scriptscriptstyle
    \dagger}_{r_2}  
  \hat{\rho}_t^{\scriptscriptstyle 0} )
  \\
  &&
  \approx 
 \sum_{r}
  \textrm{Tr}
  ( {\hat a}_{r_1}
  \hat{E}^{\bf{\scriptscriptstyle A}} (S) {\hat a}^{\scriptscriptstyle
    \dagger}_{r}  
   {\hat a}_{r}
  \hat{E}^{\bf{\scriptscriptstyle A}} (S) {\hat a}^{\scriptscriptstyle
    \dagger}_{r_2}  
  \hat{\rho}_t^{\scriptscriptstyle 0} )
  \\
  &&=
  \textrm{Tr}
  ( {\hat a}_{r_1}
  \hat{E}^{\bf{\scriptscriptstyle A}} (S) \hat{N}_{M}
  {\hat a}^{\scriptscriptstyle \dagger}_{r_2}  
\hat{\rho}_t^{\scriptscriptstyle 0} )
  =
  \textrm{Tr}
  ( {\hat a}_{r_1}
  \hat{E}^{\bf{\scriptscriptstyle A}} (S) 
  {\hat a}^{\scriptscriptstyle \dagger}_{r_2} 
    \hat{\rho}_t^{\scriptscriptstyle 0} ) .
\end{eqnarray*}
Then $ \hat{F}_t^{\scriptscriptstyle
  (1)}={({\hat{F}_t^{\scriptscriptstyle (1)}})}{}^2$ turns out to be the 
spectral measure of a self-adjoint operator in ${\cal
  H}^{\scriptscriptstyle M}$,
representing an observable ${A}_t^{\scriptscriptstyle (1)}$ 
of the microsystem.
Let us stress that in this construction an explicit time dependence of 
${F}_t^{\scriptscriptstyle (1)}$ (${A}_t^{\scriptscriptstyle (1)}$ in
the more particular case) arises in a  
quite natural way, since in  addition to the microchannel a
macrosystem dependent dynamics cannot be in general avoided. However
right  now the concept of a \emph{good detecting part} inside the system
can be easily formulated assuming that in the relevant time interval
$[t_0,t_1]$ the explicit time dependence of ${F}_t^{\scriptscriptstyle (1)}$
(${A}_t^{\scriptscriptstyle (1)}$) is either negligible or well-known
on the basis of 
macroscopic phenomenology. We shall simply forget this time dependence 
setting $\hat{F}_t^{\scriptscriptstyle (1)} (S)\approx
\hat{F}_{t_0}^{\scriptscriptstyle (1)} (S)\approx 
\hat{F}^{\scriptscriptstyle (1)} (S) $ ($\hat{A}_t^{\scriptscriptstyle
  (1)}\approx \hat{A}_{t_0}^{\scriptscriptstyle (1)} \approx 
\hat{A}^{\scriptscriptstyle (1)}  $).
Then r.h.s. of (\ref{eq:2.9}) becomes the basic formula for
probability distribution of an observable given in general by a
p.o.v.\ measure or by a self-adjoint operator ${A}^{\scriptscriptstyle
  (1)}$ in a more 
idealized situation, related to a microsystem associated with the
statistical operator $W^{\scriptscriptstyle (1)} (t)$ and produced,
living and detected 
inside the macrosystem: ${\cal
  H}^{\scriptscriptstyle M}$ is its Hilbert space and  
$  H^{\scriptscriptstyle (1)}=\sum_{r\in M}|r\rangle W_{r}\langle r|$ its
Hamiltonian. In this neat picture there is however a fundamental flaw: 
interaction with $M^{\scriptscriptstyle C}$ 
modes has been neglected. Experimental
particle physics shows us that this is indeed allowed when
experimental physicists have been clever enough, but what we have
described can never be more than an approximation. Corrections to this 
picture can be calculated: when they are small enough to preserve the
basic picture, the concept of a microsystem undergoing an unavoidable
decoherence arises in a very natural way. Let us take a statistical
operator $\hat{\rho}_{t_0}^{\scriptscriptstyle 0}$ of the form (\ref{eq:2.4})
\begin{eqnarray*}
  \lefteqn{\hat{\rho}_t = (1-\lambda)\hat{\rho}_t^{\scriptscriptstyle
      0} }
  \\
  &&+ \lambda \!\! 
  \sum_{r,r'\in M} w_{rr'} (t_0)  
  e^{-\frac{i}{\hbar}\hat{H} (t-t_0)}
  {\hat a}^{\scriptscriptstyle
    \dagger}_r e^{+\frac{i}{\hbar}\hat{H} (t-t_0)}
  \hat{\rho}_t^{\scriptscriptstyle 0} 
    e^{-\frac{i}{\hbar}\hat{H} (t-t_0)}
    {\hat a}_{r'}e^{+\frac{i}{\hbar}\hat{H} (t-t_0)} ,
 \end{eqnarray*}
by the assumption of a \emph{good detecting part} we can replace
$\hat{\rho}_t^{\scriptscriptstyle 0}$ with  
$\hat{\rho}_{t_0}^{\scriptscriptstyle 0}$ in the second term. To
take interaction with ${M^{\scriptscriptstyle C}}$ 
modes into account a suitable time scale
$\tau\approx \frac{\hbar}{\Delta W}$, where $\Delta W$ is the width of 
the energy band of the microchannel, must be considered and $\tau$
must be large enough to allow for a treatment of $M$, 
${M^{\scriptscriptstyle C}}$
interaction by a formalism similar to scattering theory, in which
states are replaced by operators, the Hamiltonian by the Liouvillean
${\cal H}= - \frac{i}{\hbar}[{\hat H},\cdot]$ and the scattering operator
by a scattering map ${\cal T}(z)$. In fact setting ${\cal H}={\cal
  H}_0 + {\cal V}$, with ${\cal H}_0={i \over \hbar}
[\hat{H}_0,\cdot]$, $\hat{H}_0 = \sum_r W_r \hat a^{\scriptscriptstyle
  \dagger}_{r}\hat a_r$ one has:
\begin{eqnarray*}
   \lefteqn{ e^{-\frac{i}{\hbar}\hat{H} (t-t_0)}
  {\hat a}^{\scriptscriptstyle
    \dagger}_r e^{+\frac{i}{\hbar}\hat{H} (t-t_0)}
    =
          {\int_{-i\infty+\eta}^{+i\infty +  \eta}}{
        dz  
        \over  
            2\pi i  
        }       \,   e^{z \tau}  
                {{  
        \left(  
        {{ z - {\cal H}}}
        \right)  
        }^{-1}}  
          {\hat a}^{\scriptscriptstyle
    \dagger}_r}
    \\
  &&=
   e^{-\frac{i}{\hbar}W_r (t-t_0)} {\hat a}^{\scriptscriptstyle
    \dagger}_r +
          {\int_{-i\infty+\eta}^{+i\infty +  \eta}}{
        dz  
        \over  
            2\pi i  
        }       \,   e^{z \tau}  
            \left[{{  
            \left( 
        {{ z - {\cal H}_0}}
        \right) 
        }^{-1}}  
        {\cal T}(z){{  
        \left(  
        {{ z - {\cal H}_0}}
        \right)  
        }^{-1}}  \right]
          {\hat a}^{\scriptscriptstyle \dagger}_r
\end{eqnarray*}
and similarly for the adjoint operator. The part depending on ${\cal
  T}(z)$ is responsible for decoherence. If
$\hat{\rho}_{t_0}^{\scriptscriptstyle 0}$ is an  
equilibrium state the treatment of this part gives in a perspicuous
way the theory of Brownian motion~\cite{art3-art4} and in the limit of 
small momentum transfers the typical dynamics of a particle undergoing 
friction and position and momentum diffusion is found. One can expect
that also in the case of a non-equilibrium
$\hat{\rho}_{t_0}^{\scriptscriptstyle 0}$ of the kind that will be
discussed in \S~\ref{s3} a
similar approach can be fruitful.
\section{Embedding of microchannel in the dynamics of a macrosystem} 
\label{s3}
\setcounter{equation}{0}
In our approach microsystems are derived entities and are no longer
the basic elementary starting point of the physical description: then 
this description must stand on its own legs by a suitable
reformulation of quantum mechanics of finite isolated non-equilibrium
system. Let us briefly recall some main points about this 
general description of macrosystems~\cite{torun}. 
The very claim that a physical
system is isolated implies the choice of a subset of observables that are
\emph{under control} by a suitable preparation procedure performed on the 
system during a preparation time interval $[T,t_0]$.
These observables are suitable slowly varying quantities, typically
densities of conserved charges ${\hat A}_j({\mbox{{\boldmath$\xi$}}})$ 
related to symmetry properties of the underlying local field
theoretical structure. By their expectations $\{
\langle
{\hat A}_j({\mbox{{\boldmath$\xi$}}})
\rangle_t
\}_j$
a set of classical fields $\{ {\zeta}_j({\mbox{{\boldmath$\xi$}}},t) \}_j$ is
determined when these expectations are reproduced by means of a
maximal von Neumann entropy state ${\hat w}[{\zeta{(t)}}]$. 
This is a generalized Gibbs state,
induced at any time $t$ by the statistical operator ${\hat \varrho}_t$ 
via the expectations $
\langle
{\hat A}_j({\mbox{{\boldmath$\xi$}}})
\rangle_t
={\mbox{{\rm Tr}}} \,
(
{\hat A}_j({\mbox{{\boldmath$\xi$}}})
{\hat \varrho}_t
)$. It depends on the operators ${\hat
  A}_j({\mbox{{\boldmath$\xi$}}})$ and the fields
${\zeta}_j({\mbox{{\boldmath$\xi$}}},t)$ and provides an entropy for
the classical state $\{ {\zeta}_j({\mbox{{\boldmath$\xi$}}},t) \}_j$.
Such classical state, though related to statistical properties of the
system that has been prepared in the time interval $[T,t]$, is taken
as an objective property of the system at time $t$. This is already
done, perhaps without complete awareness, when a velocity, a
temperature or a chemical potential field is associated to a massive
continuum. The statistical operator ${\hat \varrho}_t$, representing
preparation until time $t\geq t_0$, shows the spontaneous dynamics of the
isolated system in the time interval $[t_0,t]$, given by
$
        {\hat \varrho}_t
        =
        e^{-{{
        i
        \over
         \hbar
        }}{\hat H}(t-t_0)}
        {\hat \varrho}_{t_0}
        e^{+{{
        i
        \over
         \hbar
        }}{\hat H}(t-t_0)}
$, and can be written in the form
\begin{equation}
  \label{eq:3.1}
    {\hat \varrho}_t =      \exp
         \bigl\{
        {
        -\zeta_0(t) {\hat {\bf 1}}
        -{
        \sum_j
        \int d {\mbox{{\boldmath$\xi$}}} \,
        {\zeta}_j({\mbox{{\boldmath$\xi$}}},t)
        {\hat A}_j({\mbox{{\boldmath$\xi$}}})
        }}  
        + \int_{T}^{t} dt'\,
        {\hat {\cal S}}_t(t')   
         \bigr\}
\end{equation}
where the history part ${\hat {\cal S}}_t(t')$ for $t'\in[T,t_0]$
describes the preparation procedure and for $t'\in[t_0,t]$ can be
simply given in terms of state variables ${{\zeta}}_j
({\mbox{{\boldmath$\xi$}}},{t'})$, ${\dot{\zeta}}_j
({\mbox{{\boldmath$\xi$}}},{t'})$ and the related density and current
operators given at time $- (t-t')$ in Heisenberg picture. Since the
first term alone in the exponent already exactly gives 
$
        \langle
        {\hat
  A}_j({\mbox{{\boldmath$\xi$}}})
        \rangle_t
$, an expansion with respect to the history term becomes very natural
and e.g., at first order leads to an evolution equation
\begin{equation}
  \label{eq:3.2}
  \frac{d}{dt}
  \langle
{\hat A}_j({\mbox{{\boldmath$\xi$}}})
\rangle_t
=
        {\mbox{{\rm Tr}}} \,
        (
        \dot{{\hat A}_j}({\mbox{{\boldmath$\xi$}}})
        {\hat w}[{\zeta{(t)}}]
        )
        +
         \int_T^t dt'\,
       \mbox{\boldmath $\langle$}
        \dot{{\hat A}_j}({\mbox{{\boldmath$\xi$}}})
        ,
        {\hat {\cal S}}_t(t')
        \mbox{\boldmath $\rangle$}_{{\hat w}[{\zeta{(t)}}]}       
        + \ldots
\end{equation}
where ${\hat w}[{\zeta{(t)}}]$ is the generalized Gibbs state
associated to the classical state at time $t$. To the expectation
values of the operators $\dot{{\hat A}_j}({\mbox{{\boldmath$\xi$}}})$
calculated with ${\hat w}[{\zeta{(t)}}]$ corrections responsible for
irreversibility arise by the history term, which brings in foreground 
an integral over $t'$ of the two point Kubo correlation functions
between operators $\dot{{\hat A}_j}({\mbox{{\boldmath$\xi$}}})$ and
operators $\dot{{\hat A}_j}({\mbox{{\boldmath$\xi$}}}',- (t-t'))$,
${{\hat A}_j}({\mbox{{\boldmath$\xi$}}}',- (t-t'))$ in the macrostate
${\hat w}[{\zeta{(t)}}]$. Now the possibility of a great
simplification imposes on our attention: as at equilibrium, these
correlation functions, at least inside a time integral with well
shaped classical state parameters, could practically vanish if
$t'<t-\tau$, $\tau$ being a characteristic decay time; then $ \int_T^t
dt' \rightarrow \int_{t-\tau}^t
dt' $, thus eliminating memory of the preparation procedure for
$t>t_0+\tau$ and memory of previous classical state if it variates
slowly enough during a time interval $\tau$. We call such a situation 
\emph{simple
dynamics}: it dominates a large part of equilibrium
thermodynamics. 
However we also discover a large arena where a behavior different from
\emph{simple dynamics} can arise. One expects that when the fields
${\zeta}_j({\mbox{{\boldmath$\xi$}}},t_1)$ are inhomogeneous enough
around time $t_1$,
depletion of certain modes can arise:
${\hat a}_r{\hat w}[{\zeta{(t_1)}}]\approx 0$ if $r\in M$, then the
part of previous history related to creation of these modes might
present a slowly decaying contribution. Let us write:
\begin{equation}
  \label{eq:3.3}
  {\hat {\cal S}}_{t_1}(t')  = 
{\hat {\cal S}}^{\scriptscriptstyle (S)}_{t_1}(t') + 
{\hat {\cal S}}^{\scriptscriptstyle (M)}_{t_1}(t')  
\end{equation}
where $ {\hat {\cal S}}^{\scriptscriptstyle (S)}_{t_1}(t') $ 
does not create particles in
the $M$ modes, thus yielding through (\ref{eq:3.1}) (with $ {\hat {\cal
    S}}^{\scriptscriptstyle (S)}_{t_1}(t') $ at place of 
$ {\hat {\cal S}}_{t}(t') $) a
statistical operator ${\hat \varrho}_{t_1}^{\scriptscriptstyle (S)}$
with 
\emph{simple
  dynamics}, while the full statistical operator ${\hat
  \varrho}_{t_1}$ can be written by an expansion  with respect to $
{\hat {\cal S}}^{\scriptscriptstyle (M)}_{t_1}(t') $, preserving positivity:
\begin{equation}
  \label{eq:3.4}
  {\hat \varrho}_{t_1} = \lambda 
          \left[
        {\hat {\bf 1}}
        +
        \int_{t_1 - \tau}^{t_1} dt'\,
        {\hat {\mbox{{S}}}}^{\scriptscriptstyle (M)}_{t_1}(t')
        \right]
        {\hat \varrho}_{t_1}^{\scriptscriptstyle (S)}
        \left[
        {\hat {\bf 1}}
        +
        \int_{t_1 - \tau}^{t_1} dt'\,
{\hat {\mbox{{S}}}}^{\scriptscriptstyle 
(M)}_{{\bar t}}{}^{\scriptscriptstyle\dagger}(t')
        \right],
\end{equation}
where ${\hat {\mbox{{S}}}}^{\scriptscriptstyle (M)}_{t_1}(t')$ 
is essentially
determined by ${\hat {\cal S}}^{\scriptscriptstyle (M)}_{t_1}(t')$. 
Let us write:
\begin{displaymath}
           \int_{t_1 - \tau}^{t_1} dt'\,
        {\hat {\mbox{{S}}}}^{\scriptscriptstyle (M)}_{t_1}(t')
        =
        \sum_{r\in M} {\hat a}^{\scriptscriptstyle \dagger}_r
                   \int_{t_1 - \tau}^{t_1} dt'\,
        \int_{\omega_s} d {\mbox{{\boldmath$\xi$}}'} \,
        {\hat A}_r(- (t_1 - t'), {\mbox{{\boldmath$\xi$}}'})
        ,
\end{displaymath}
where the field operator ${\hat A}_r(- (t_1 - t'),
{\mbox{{\boldmath$\xi$}}'})$ acts as annihilation operator typically
for ${\mbox{{\boldmath$\xi$}}'}$ inside some space region $\omega_s$.
If a local observable ${\hat B}({\mbox{{\boldmath$\xi$}}},t)$ is
considered at time $t$, such that correlations between the space-time
point $({\mbox{{\boldmath$\xi$}}},t)$ and region $\omega_s$ are
negligible, one can write
\begin{eqnarray*} 
\lefteqn{  {\mbox{{\rm Tr}}} \,
   [
   {\hat B}({\mbox{{\boldmath$\xi$}}},t) \!\!
 \sum_{r,r'\in M} 
  {\hat a}^{\scriptscriptstyle
    \dagger}_r 
                    \int_{t_1 - \tau}^{t_1} \!\!\! dt' \!\!
        \int_{\omega_s} \!\!\! d {\mbox{{\boldmath$\xi$}}'} \,
        {\hat A}_r(- (t_1 - t'), {\mbox{{\boldmath$\xi$}}'})
        {\hat \varrho}_{t_1}^{\scriptscriptstyle (S)}
} \cdot
\\
&&
\cdot
                    \int_{t_1 - \tau}^{t_1} \!\!\! dt' \!\!
        \int_{\omega_s} \!\!\! d {\mbox{{\boldmath$\xi$}}'} \,
        {\hat A}^{\scriptscriptstyle
    \dagger}_r(- (t_1 - t'), {\mbox{{\boldmath$\xi$}}'})
  {\hat a}_r 
   ]
   \approx
    {\mbox{{\rm Tr}}} \,
   [
   {\hat B}({\mbox{{\boldmath$\xi$}}},t)\!\!
 \sum_{r,r'\in M} 
  {\hat a}^{\scriptscriptstyle
    \dagger}_r 
        {\hat \varrho}_{t_1}^{\scriptscriptstyle (S)}
  {\hat a}_r 
   ]
   \sigma_{r' r}
\end{eqnarray*}
with
\begin{displaymath}
  \label{eq:3.5}
    \sigma_{r' r}
    =
      {\mbox{{\rm Tr}}} \,
   [
                     \int_{t_1 - \tau}^{t_1} \!\!\! dt' \!\!
        \int_{\omega_s} \!\!\! d {\mbox{{\boldmath$\xi$}}'} \,
        {\hat A}_r(- (t_1 - t'), {\mbox{{\boldmath$\xi$}}'})
        {\hat \varrho}_{t_1}^{\scriptscriptstyle (S)}
                    \int_{t_1 - \tau}^{t_1} \!\!\! dt' \!\!
        \int_{\omega_s} \!\!\! d {\mbox{{\boldmath$\xi$}}'} \,
        {\hat A}^{\scriptscriptstyle
    \dagger}_r(- (t_1 - t'), {\mbox{{\boldmath$\xi$}}'})
   ] ,
\end{displaymath}
being a positive matrix describing the way in which the normal modes of
$M$ are feeded by destruction of particles in $\omega_s$: in this way
we are recovering the starting point of \S~\ref{s2}.
\begin{acknowledgments}
L.~L.\ and B.~V.\ gratefully acknowledge financial support by MURST
under Cofinanziamento. B.~V. also acknowledges support by MURST under
Progetto Giovani.  
\end{acknowledgments}

\end{document}